\documentstyle[preprint,aps,eqsecnum]{revtex}

\begin{document}
\draft

\title{Dynamics of Dissipative Binary Collisions}

\author{G.Fabbri${}^{1,2}$, M.Colonna${}^{1}$, and M.Di Toro${}^{1}$}

\address{${}^{1}$)Laboratorio Nazionale del Sud, Via S. Sofia 44,
I-95123, Catania, Italy}

\address{${}^{2}$)Dipartimento di Fisica, Universit\`a di
Firenze, and INFN,\\ Largo E.Fermi 2, I-50125, Firenze, Italy}

\maketitle
\begin{abstract}
In this paper we discuss the reaction mechanisms that occur in the
overlap zone for 
semi-peripheral heavy ion collisions at intermediate energies.
In particular we focus on the development of neck instabilities, 
that could determine 
a possible increase of dynamical fluctuations. 
As observed in recent experimental data, 
at beam energies just above $10~MeV/A$ the most relevant 
expected consequence is the
possibility to obtain large variances in the projectile-like and target-like
observables.
With increasing beam energy we pass to a mid-rapidity fragment production. In
this way we predict a smooth transition from a deep-inelastic to a
fragmentation reaction mechanism. 
\end{abstract}


\section{Introduction}

In recent years the study of the reaction mechanisms involved in heavy ion
collisions at intermediate energies has been the subject of several
experimental and theoretical investigations all around the world. 
In particular, recent experimental results have raised the attention on the 
possibility to reveal prompt intermediate mass fragment
(IMF) production, as well as large fluctuations in the
observables related to projectile-like (PLF) and target-like (TLF)
fragments  in semi-peripheral collisions \cite{exp}.  

In this paper we will concentrate on the dissipation mechanisms occurring
in medium momentum-transfer reactions, corresponding
to semi-peripheral impact parameters, at beam energies slightly above 
$10~MeV/A$. 
In this energy range, interesting results on the excitation energy sharing 
between PLF and TLF fragments and on the observation of large mass variances 
have been recently reported in Ref.\cite{Casini}.

We will investigate the dynamics of the nuclear overlap
zone (the "neck" region): 
the occurrence of volume instabilities in this zone 
helps the break-up of the dinuclear 
system formed in the earlier stage of the reaction, leading to two primary
fragments in the exit channel. 
Volume instabilities can develop 
due to the coupling between stochastic nucleon-nucleon collisions 
and the nucleon exchange process
already present also at lower energies and extensively studied
in deep inelastic collisions and fusion-fission events, from both
theoretical and experimental points of view 
\cite{Sierk,D_E_Fields,Olmi}.
We will discuss the relative importance of these two sources of dissipation
and fluctuations in the energy range considered. In particular, 
the reaction that we will consider here, $^{100}$Mo + $^{120}$Sn at $14~MeV/A$, 
lies in the 
beam energy region where two-body collisions have just started to play a role
and, consequently, as we will see in the following, the dissipation mechanism
is mostly determined by the one-body nucleon exchange.  
According to the calculations presented below, at this energy 
fluctuations due to two-body collisions just start to 
be important.

The detectable consequence of the occurrence of instabilities should be 
a clear increase 
of the variances of all observables (masses, charges, angles, velocities,
angular momenta...) of projectile-like (PLF) and target-like (TLF) fragments
and, at higher energies, 
the possibility of IMF formation from the neck region.

\section{Dynamical Evolution}

The main purpose  of this 
paper is to study the mechanisms which occur when the di-nuclear system, 
formed in the earlier stage of the reaction, breaks up into pieces. 
The system will easily split apart if, after the shock and the initial 
compression, the rarefaction phase leads  the density in the 
overlap zone below the critical density \cite{Bertsch}. 
The occurrence of volume instabilities can therefore explain the break-up
of the system in two primary fragments (like in deep-inelastic collisions),
or even determine a prompt IMF emission from the neck region 
at higher bombarding energies, where the 
nucleon-nucleon collision rate becomes more important \cite{npa}.

It is well known that, when instabilities are present, fluctuations 
become extremely important. In fact, in unstable situations, fluctuations 
are usually amplified and may lead the system towards various
patterns that are different from the one associated with the mean trajectory 
behaviour. In the following we will try to relate the presence of volume 
instabilities to the possibility, due to the growth of fluctuations, to 
obtain large variances for the observables associated with 
projectile-like (PLF) and target-like (TLF) fragments in semi-peripheral 
reactions at around $15~MeV/A$. 

A parameter of crucial importance is the time interval during which the 
di-nuclear system interacts and exchanges nucleons before its 
break-up. This time essentially depends on impact 
parameter and beam energy.  If it 
is long compared to the characteristic time for 
the growing of spinodal instabilities ($\tau \approx 
100-200~fm/c$) \cite{CC},
the effect of the enhancement of 
fluctuations due to instabilities will be averaged out by the 
mean-field propagation; in this case we 
expect to observe just the equilibrium fluctuations associated with the 
stochastic nature of nucleon exchange and/or nucleon-nucleon 
collisions in stable systems \cite{npa,equili}. 
This is for instance the case of deep-inelastic collisions at low energy,
below $10~MeV/A$.  This corresponds mostly to the situation where, in average,
PLF and TFL have the same temperature, i.e. the total excitation energy is
shared according to the mass ratio of the two spectators.

On the other hand, if the interaction time is of the same order of 
magnitude of the instability growth time, these fluctuations will be 
amplified and will lead to variances larger than those expected on the basis
of statistical equilibrium, for all observables related 
to the primary products of the reaction.  
This is extremely interesting 
since recent experimental results have already raised the attention on 
the possibility to obtain a big variety of masses of PLF and TLF, as well 
as IMF emission from the neck region in 
semi-peripheral heavy ion collisions at intermediate energies 
\cite{exp}, where the condition on the interaction time discussed 
before surely applies.

\section{ Theoretical framework}
In order to have a dynamical description that incorporates fluctuations, we will
perform calculations considering a stochastic mean field approach.  
In such a kind of theories the nuclear 
system is still described by its one body density function in 
phase space $f({\bf r}, {\bf p}, t)$, while this function may experience 
a stochastic evolution in response to the action of a fluctuating
source term, in some analogy with the Brownian motion.

Recently kinetic one-body equations of the Boltzmann-Nordheim-Vlasov 
(BNV) (or Boltzmann-Uehling-Uhlenbeck (BUU)) type have been extended by 
the introduction of a fluctuating term, coming from considerations associated 
with the random nature of 
the nucleon-nucleon collision integral \cite{Ayik,Randrup,Phil}.
The resulting Boltzmann-Langevin (BL) 
equation reads:
\begin{equation}
{{\partial f}\over{\partial t}}
+({\bf v}\cdot{{\partial f}\over{\partial {\bf r}}}
-{{\partial U}\over{\partial {\bf r}}}\cdot{{\partial f}\over{\partial {\bf 
p}}}) =
\bar{I}[f]+\delta I[f]
\end{equation}
where $U[f]$ is the self-consistent mean field potential,
$\bar{I}$ represents the average effect of the 
collisions,
while $\delta I$ denotes the fluctuating remainder
(the Langevin term).

In our calculations we will use a simplified approach: once local volume
instabilities are encountered, we implement in the code local fluctuations of
the density according to the amplitude predicted by the BL theory at the 
temperature and density considered \cite{Ayik}.  This procedure is described in
details in Ref.\cite{Alfio}. 
These fluctuations are then amplified and lead 
to several situations of mass and excitation energy exchange in the exit
channel. 

The BNV equation is solved within the test particle method \cite{Aldo},
using the code TWINGO \cite{Alfio_twingo}. 
The following mean-field parameterization has been considered:
\begin{equation}
U(\rho) = A~(\rho/\rho_0) + B~(\rho/\rho_0)^\sigma,
\end{equation}
with $A = -356~MeV$, $B = 303~MeV$, $\sigma = 7/6$, that gives a "soft"
equation of state, with a compressibility modulus $K = 200~MeV$.  
$\rho_{0}$ is the equilibrium density of symmetric nuclear matter. 
We have checked that the numerical fluctuations introduced in this way are
negligible  when compared to the physical fluctuation amplitude that we
implement when instabilities are encountered. 
\vskip 0.2truecm

\section{Description of results}

Our aim is mainly to compare the results on the energy sharing 
between PLF and TLF with the 
experimental findings obtained recently at GSI \cite{Casini} for the 
reaction $^{100}$Mo + $^{120}$Sn at $14~MeV/A$. Because of intrinsic limits
of the apparatus, in those experiments observables related to PLF
were evaluated in direct and reverse kinematics. We have investigated 
the evolution of the BNV dynamics from semi-peripheral to peripheral
collisions. 

At impact parameters less than $b = 7~fm$ the system follows the 
average path of 
incomplete fusion, converting the incoming energy partly into rotational 
motion of the di-nuclear system and partly into 
internal excitation energy. Increasing the impact parameter the average 
trajectory of the system evolves towards deep-inelastic process configurations.
We observe a binary mechanism, that preserves the identity of the two 
colliding nuclei, but
converting a quite large fraction of the available energy and angular momentum
into thermal energy and intrinsic spin of the primary fragments.
As mentioned above, we are mainly interested in evaluating 
the energy sharing between the two partners.
Two situations are usually indicated: The quasi-elastic limit and the thermal 
limit. In the quasi-elastic 
case the partners come in contact for a relatively short time, allowing
only few nucleons to pass from one nucleus to the other.
Concerning the excitation energy sharing, this leads in average to the 
equipartition between the two primary fragments.
This result is found in general models, as the Nucleon Exchange Model (NEM) 
\cite{NEM}, where, due to the stochastic nature of the nucleon exchange
process,  
the two partners have fluctuating masses, but essentially the same
excitation energy. 

On the other side, when the collision time is long enough, the two
reaction partners can exchange a very large number of nucleons and the system goes
towards thermalization. As a consequence, the excitation energy is divided
between the fragments proportionally to their masses. This is what
is usually called the thermal limit.

The energy dissipated in the reaction is given essentially by the total 
kinetic energy loss ($TKEL$), defined as the difference, in the center-of-mass 
reference frame, between the 
initial available kinetic energy $E$ and the total kinetic energy ($TKE$) 
of the two primary fragments in the exit channel:
\begin{equation}
TKEL = E - TKE. 
\end{equation} 
In the calculation of $TKEL$ we have
taken into account the Coulomb repulsion between the two primary fragments in 
the final stage.
The $TKEL$ strongly depends on the impact parameter and the 
interaction time (the time after which the fragments separate), 
as we show in Table I. 

Because of this, 
from the experimental point of view, $TKEL$ represents a good parameter 
to explore the evolution of the system between the two limits of 
reaction mechanisms indicated above.

\subsection{Calculations including fluctuations}

In order to perform a comparison with the experimental data \cite{Casini},
we study the correlations that can arise
between the net-mass transfer and the excitation energy of PLF and TLF.
To tackle this problem we stress that 
it is necessary to perform an event-by-event analysis, since 
the fluctuations around the average dynamics play an important role.
We have concentrated our analysis on the nature of the binary processes
occurring at $b=8.5,9,10~fm$. 
In Fig.1 we show the time evolution of three events corresponding
to these values of $b$. The $TKEL$'s corresponding to these
values of the impact parameter are indicated in Table I.  
It can also be seen in the Table that 
the separation time is still quite long in the $b = 8~fm$ case 
and this complicates significantly the analysis. 
For this reason we show results starting from $b = 8.5~fm$.
For each impact parameter several events have been considered. 
The calculations show that the system in average evolves 
from the quasi-elastic to the thermal limit along with the rise of $TKEL$. 
Indeed, in the case of the more peripheral
collision ($b = 10~fm$) the excitation energy is, in average, 
equally shared between the
two fragments, while for $b=8.5~fm$ the largest fragment is more excited.
Following the entire dynamics of the system, it is possible to evaluate
how many nucleons are exchanged in the three cases. 
This average number is shown in Table II, 
along with the $TKEL$ bins, as
a function of the impact parameter. 

In the Table II are shown also the
average masses of the fragments and their variances. As expected, 
the number of exchanges increases rapidly with the
dissipation degree of the reaction. 
The variances observed are only due to
dynamical fluctuations, that develop as soon as the di-nuclear
system encounters volume or shape instabilities. 
As mentioned before, this kind of 
fluctuations is accounted for in our stochastic
simulations.  
However, it should be noticed that, 
also in stable situations, equilibrium fluctuations are present due to the
stochastic nature of the nucleon exchange process.  Because of the use of test
particles for solving the transport equation, these fluctuations are
reduced by a factor $1/N_{test}$, being $N_{test}$ the number of test particles
per nucleon. 
We incorporate also these fluctuations (that we will call
"statistical" fluctuations),
by implementing  each stochastic event calculation by 
the procedure of random clustering of the one-body distribution, 
introduced in Ref.\cite{Bonasera}. The method is widely discussed
also in Ref.\cite{equili}. Starting from a dynamical event we get
many "statistical" events, each of them constructed by randomly choosing 
a sample of $N_p=A_p-Z_p$ and $Z_p$ test particles among all test particles 
associated with 
neutrons and protons of the projectile, and similarly for the protons
and neutrons of the target. Then the primary fragments are reconstructed 
using a coalescence procedure described in Ref.\cite{Bonasera}.
In this way we reconstruct the "statistical" variances, 
that are essentially given by the average number of 
exchanged nucleons (Ref.\cite{npa}).
As it can be seen from Table II, these fluctuations are
larger than the dynamical variances.

\subsection{Comparison with experimental data}
In the experimental data (Ref.\cite{Casini}),
it has been unambiguously demonstrated in a model independent way 
that a correlation between the net mass transfer and the excitation energy of
the fragments exists even for relatively high energy dissipation. 
Moreover in that paper the authors show that, also in events with the
two primary fragments having the same mass, the excitation energy is not 
equally shared 
between the two fragments, being much more excited the one 
that gains nucleons. Such a result is in disagreement with both the
quasi-elastic limit, usually described by the NEM,  
and the thermal limit, which predict the equipartition 
of the excitation energy for equal-mass events. Hence Casini et al. 
(Ref.\cite{Casini})
advocate the existence of important dynamical effects in order to explain the
data. 

As reported in Table II, in our simulations 
we observe non-negligible fluctuations that come from dynamical effects (neck
instabilities), that are responsible for the variances indicated in the Table 
and 
many nucleon exchanges between the two reaction partners.
Once the statistical fluctuations have been implemented (according to the
procedure \cite{Bonasera} recalled above), we observe, in each dynamical event,
a broadening of the mass distribution according to the formula $\sigma_{stat}^2
 = 
{\bar n}_{exch}$.  
One can notice that the "statistical variances" are larger
that the ones given by dynamical instability effects. Indeed, at $14~MeV/A$,
dynamical instabilities have just started to play a role. 
After introducing the "statistical" fluctuations, we can now calculate the
masses and the excitation energies of the two partners event by event. 
In Fig.2 we show the excitation energy of the PLF and TLF 
fragments, 
obtained at b = 9 fm,  as a function of their masses. 
It is possible to see that, as already stressed before, in absence
of net mass transfer (i.e. for $A_{PLF} = 100$, $A_{TLF} = 120$), 
we observe the same excitation energy for the 
two reaction partners. Then we see that the fragment that receives a given
amount of nucleons from the partner is more excited, in agreement with the
trend observed in the experimental data.   However, it should be noticed that 
in our calculations, this result does not come from dynamical effects,
as suggested in Ref.\cite{Casini}, because we have shown that dynamical
variances are small compared to the statistical ones. Hence we conclude that
in our calculations this kind of correlations arises  
from statistical processes, namely
from the nucleon exchange process.

Actually, as stressed in recent publications \cite{chatto}, this feature 
can be understood 
also within the NEM. 
It is often argued that the NEM predicts equal energy sharing between 
the two primary fragments (see Ref.\cite{NEM}).  
However this is true only if the net mass drift is much smaller than the total
diffusion of nucleons. If this condition is not fulfilled a correlation 
between excitation energy and net mass transfer can arise. This is related 
essentially to the fact that in a single exchange the hole excitation 
induced by the leaving nucleon in the donor nucleus is much lower 
than the particle excitation created in the recipient nucleus. 
So an asymmetry in the fragment excitation energy 
is expected in events with a net mass transfer, being more excited the
nucleus that gains nucleons.
If the total number of exchanges is not much higher than the net transfer, 
this asymmetry is detectable.

The behaviour observed at $b = 10~fm$ is similar to what is obtained
at $9~fm$ (equipartition of excitation energy between the primary
fragments) but variances are much less.

At $b = 8.5~fm$ the system goes towards the equal temperature limit; hence
the excitation energies of the fragments result to be just proportional to 
their masses.
This trend is also present in the data, when increasing the $TKEL$,
even if the equal temperature limit is reached for higher values of
$TKEL$.

As one can see from Fig.2, 
the experimental mass variances are not correctly reproduced in the 
calculations,
since the range of masses detected as PLF and TLF is broader in the data. 
This could indicate a more important contribution of dynamical effects and
instabilities that, in our calculations, at $14~MeV/A$ just start to appear. 

We expect to find such a contribution when
rising the energy.
Indeed, calculations at higher energies ($25~MeV/A$) are presently in progress
and 
seem to suggest that dynamical instabilities play an important
role in the evolution of the system, giving a very relevant 
contribution to the  final variances. 

\section{Conclusions}
We have investigated the dynamics of semi-peripheral heavy ion collisions at
energies around $15~MeV/A$.   The interest of this kind of study lies on the fact
that in recent experimental data large fluctuations have been measured in the
observables related to PLF and TLF fragments, that cannot be explained just
on the basis of thermal "equilibrium" processes. 
The dynamical mechanisms and the possible occurrence of instabilities
have been investigated, in the case of the reaction $^{100}$Mo + $^{120}$Sn 
at $14~MeV/A$,
in the framework of a stochastic mean-field approach.  We show that
dynamical effects, related to the occurrence of volume and shape instabilities
in the neck region, determine an increase of the variances, but at $14~MeV/A$
this effect is small when compared to the broadening due to statistical
fluctuations. 
At the energy considered dynamical fluctuations just start to play a role.
On the basis of previous calculations, larger effects are predicted at higher
energy \cite{npa}, up to cluster formation in the "neck" region, with variances 
much above the statistical evaluation.  In this way we expect a quite smooth
transition in the reaction mechanism for dissipative collisions, from
deep-inelastic to fragmentation: "Natura non facit saltus" \cite{Luc}.   

\acknowledgements
We wish to thank F.Matera for valuable discussions.

\begin{table}
\caption{The average total kinetic energy loss, calculated in the BNV 
simulations, as
a function of the impact parameter. The average time needed for 
the separation of
the two primary fragments is also reported.
\label{t:1}}
\begin{tabular}{ccc}
$b~(fm)$ & $TKEL~(MeV)$ & Interaction time $(fm/c)$\\
\tableline
8 & 453 & 450\\
8.5 & 387 & 350\\
9 & 347 &  300\\
9.5 & 272& 240\\
10 & 194 & 210\\
10.5 & 138 & 180\\
11 & 93 & 150\\
\end{tabular}
\end{table}

\vskip 3cm

\begin{table}
\caption{The TKEL bin, the statistical mass variance, 
the average PLF mass and associated dynamical variance, the average TLF mass 
and 
associated dynamical variance, 
calculated in the stochastic simulations, as a function of the impact
parameter. The statistical variance is given by the average number of
exchanged nucleons (see text).
\label{t:2}}
\begin{tabular}{ccccccc}
$b~(fm)$ & $TKEL~bin~(MeV)$ & $\sigma_{stat}^2$ & 
$A_{PLF}$ &  $\sigma_{PLF}^{2}$ & $A_{TLF}$ & $\sigma_{TLF}^{2}$\\
\tableline
8.5 & 369.4--404.3 & 63.7 & 99.1 & 2.69 & 120.7 & 2.69\\    
9 & 327.2--367.3 & 50.4 & 101.5 & 4.49 & 118.4 & 4.71\\
10 & 177.9--210.7 & 22.7 & 97.9 & 1.08 & 122.1 & 1.08\\
\end{tabular}
\end{table}

\begin{figure}
\caption{Contour plots of the density in the reaction plane 
for the collision $^{100}$Mo + $^{120}$Sn at $14~MeV/A$, at
three different impact parameters: (a) $b = 8.5~fm$,
(b) $b = 9~fm$, and (c)$b=10~fm$. The size of the box is $40~fm$.
}

\label{f:1}
\end{figure}

\begin{figure}
\caption{The average number of evaporated nucleons for the collision
$^{100}$Mo + $^{120}$Sn at $14~MeV/A$ as a function of the primary mass
of the PLF fragment, in the direct (crosses) and reverse (open circles)
kinematics: (a) Experimental results for $TKEL = 300-350~MeV$,
Ref. [2]; (b) Our calculations for $b = 9~fm$, that corresponds
to $TKEL = 325-370~MeV$. The full (dashed) line fits the points obtained
in the direct (reverse) kinematics. 
}

\label{f:2}
\end{figure}  


\begin{thebibliography}{99}


\bibitem{exp} 
R. Bougault {\em et al.}, Nucl. Phys. {\bf A587}, 499 (1995);
M.F. Rivet, B. Borderie, Ch. Gr\'egoire, D. Jouan, and B. Remaud,
 Phys. Lett. B {\bf 215}, 55 (1988);
L. Sobotka, Phys. Rev. C {\bf 50}, R1272 (1994);
J.F. Lecolley {\em et al.}, Phys. Lett. B {\bf 354}, 202 (1995);
C.P. Montoya {\em et al.}, Phys. Rev. Lett. {\bf 73}, 3070 (1994);
J. T\~{o}ke {\em et al.}, Phys. Rev. Lett. {\bf 75}, 2920 (1995); 
J. T\~{o}ke {\em et al.}, Nucl. Phys. {\bf A583}, 519c (1995); 
W. Lynch, Nucl. Phys. {\bf A583}, 471c (1995);
N. Colonna {\em et al.}, Proc. of the Int. Work. on Gross Properties
	  of Nuclei and Nuclear Excitations, Hirschegg 1994, Ed. H.Feldmeier. 



\bibitem{Casini} G. Casini  {\em et al.}, Phys. Rev. Lett. {\bf 78}, 828 (1997).

\bibitem{Sierk} A.J. Sierk, Phys. Rev. C {\bf 33}, 2039 (1986). 

\bibitem{D_E_Fields} D.E. Fields, K. Kwiatkowski, K.B. Morley, E.
Renshaw, J.L. Wile, S.J. Yennello, V.E. Viola, and  R.G. Korteling,
 Phys. Rev. Lett. {\bf 69}, 3713 (1992).

\bibitem{Olmi} A. Olmi, U. Lynen, J.B. Natowitz, M. Dakowski,
 P. Doll, A. Gobbi, H. Sann, H. Stelzer, R. Bock, and D. Pelte,
 Phys. Rev. Lett. {\bf 44}, 383 (1980). 

\bibitem{Bertsch} G. Bertsch and D. Mundinger, Phys. Rev. C {\bf 17}, 1646
 (1978). 

\bibitem{npa} M. Colonna, M. Di Toro, and A. Guarnera, Nucl. Phys. {\bf A589},
 160 (1995). 

\bibitem{CC} M. Colonna and Ph. Chomaz, Phys. Rev. C {\bf 49}, 1908 (1994). 

\bibitem{equili} A. Bonasera, M. Colonna, M. Di Toro, F. Gulminelli, and
A. Smerzi, Nucl. Phys. {\bf A572}, 171 (1994).

\bibitem{Ayik} S. Ayik and Ch. Gr\'egoire, Phys. Lett. B {\bf 212}, 269 (1988).

\bibitem{Randrup} 
J. Randrup and B. Remaud, Nucl. Phys. {\bf A514}, 339 (1990). 

\bibitem{Phil}
Ph. Chomaz, G.F. Burgio, and J. Randrup, Phys. Lett. B {\bf 254}, 340 (1991).

\bibitem{Alfio}
A. Guarnera, M. Colonna, and Ph. Chomaz, Phys. Lett. B {\bf 373}, 267
 (1996).  


\bibitem{Aldo}
A. Bonasera, F. Gulminelli, and J. Molitoris, Phys. Rep. {\bf 243},
 1 (1994).  

\bibitem{Alfio_twingo}
A. Guarnera, Ph.D. Thesis, Caen (1996). 

\bibitem{NEM} J. Randrup, Nucl. Phys. {\bf A307}, 319 (1978);
 Nucl. Phys. {\bf A327}, 490 (1979);
 W.U. Schr{\"o}der, J.R. Huizenga, in {\em Treatise in Heavy Ion Science},
 edited by D. Bromley (Plenum, New York, 1984) p.115;
 J. Randrup and R. Vandenbosch, Nucl. Phys. {\bf A474}, 219 (1987).

\bibitem{Bonasera}
Ch. Gr\'egoire and M. Zielinska Pfab\'e, Phys. Rev. C {\bf 37},
 2594 (1988);
A. Bonasera, M. Colonna, M. Di Toro, F. Gulminelli, and H.H. Wolter,
   Phys. Lett. B {\bf 244}, 169 (1990).


\bibitem{chatto} S. Chattopadhyay and D. Pal, Phys. Rev. C {\bf 42},
 R2283 (1990). 

\bibitem{Luc} "Nature does not make jumps", T.C. Lucretius, "De Rerum Natura",
60 B.C..

\end{thebibliography}
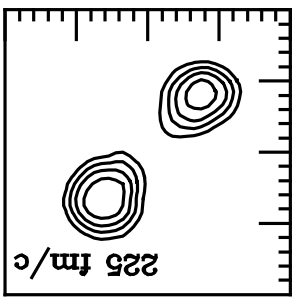
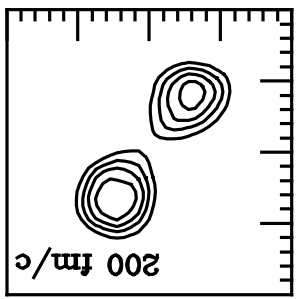
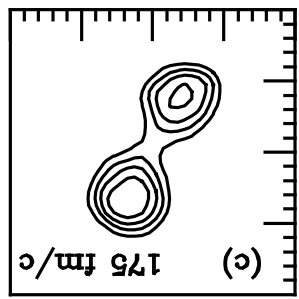
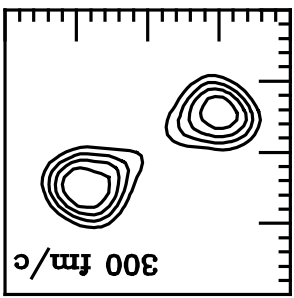
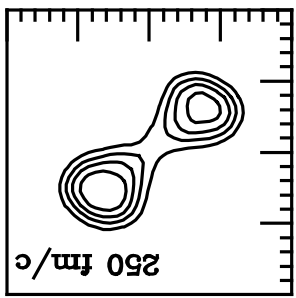
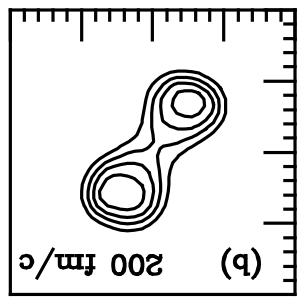
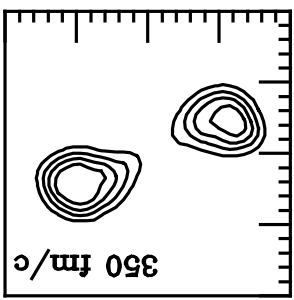
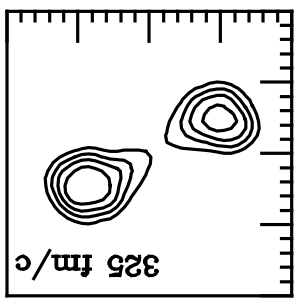
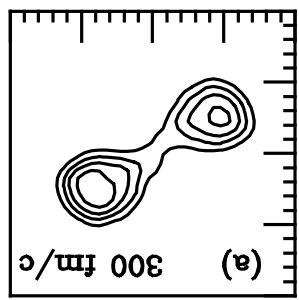
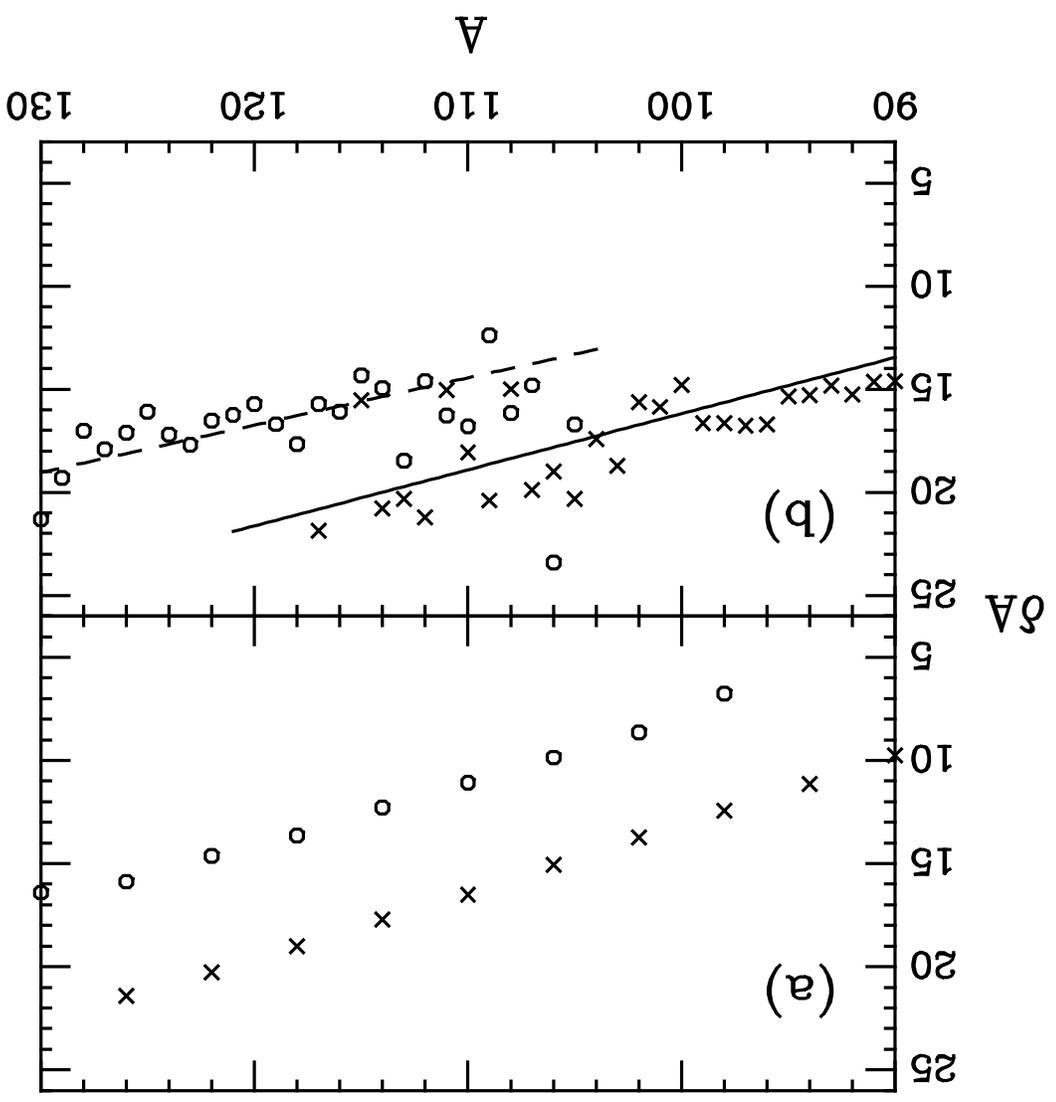
\end{document}